\documentclass[amsmath,aps,prl,showpacs,psfig,epsfig,twocolumn]{revtex4}
\usepackage{graphicx}
\input{epsf}

\begin{document}
\draft
\title{Quantum escape of the phase in a strongly driven Josephson junction}

\author{M.~V. Fistul}
\affiliation{Physikalisches Institut III, Universit{\"a}t
Erlangen-N{\"u}rnberg, D-91058 Erlangen, Germany}
\author{A. Wallraff}
\altaffiliation[Current address: ]{Department of Applied Physics,
Yale University, New Haven, CT 06520, USA}
\affiliation{Physikalisches Institut III, Universit{\"a}t
Erlangen-N{\"u}rnberg, D-91058 Erlangen, Germany}
\author{A.~V. Ustinov}
\affiliation{Physikalisches Institut III, Universit{\"a}t
Erlangen-N{\"u}rnberg, D-91058 Erlangen, Germany}

\date{\today}
\begin{abstract}
A quantum mechanical analysis of the Josephson phase escape in the
presence of both dc and ac bias currents is presented. We find
that the potential barrier for the escape of the phase is
effectively suppressed as the resonant condition occurs, i.e.~when
the frequency $\omega$ of the ac bias matches the Josephson
junction energy level separation. This effect manifests itself by
a pronounced drop in the dependence of the switching current $I_s$
on the power $W$ of the applied microwave radiation and by a
peculiar double-peak structure in the switching current
distribution $P(I_s)$. The developed theory is in a good accord
with an experiment which we also report in this paper. The
obtained features can be used to characterize certain aspects of
the quantum-mechanical behavior of the Josephson phase, such as
the energy level quantization, the Rabi frequency of coherent
oscillations and the effect of damping.
\end{abstract}


\maketitle
Great attention has been devoted to the experimental and
theoretical study of Josephson junctions in the presence of
externally applied microwave radiation. The well known effects are
Shapiro steps in the current-voltage characteristics, and features
due to photon-assisted tunneling of quasiparticles \cite{Tinkham}.
Moreover, microwave radiation can be used to probe the {\it
macroscopic quantum-mechanical} behavior of the Josephson phase
such as the energy level quantization
\cite{Clarke1,Clarke2,Larkin} and the recently observed quantum
coherent Rabi oscillations \cite{Han,Martinis}. The observation of
these effects has proven that the laws of quantum mechanics can be
applied to dissipative systems. The interest in this field has
also been boosted by the possibility to use Josephson junctions as
superconducting qubits \cite{Han,Martinis}.

The quantum mechanical properties of the Josephson phase can be
studied most directly when the Josephson phase interacts
resonantly with the external microwave radiation
\cite{Clarke1,Clarke2,Han,Martinis,WalUst}. Such an interaction
leads to the resonant absorption of photons, which induces
transitions between the energy levels of the Josephson phase and
thus to a pronounced increase of the population of excited levels.

It is also well known that thermal and quantum fluctuations lead
to a premature \emph{switching} of the junction from the
superconducting state to the resistive state at a random value of
dc bias $I_s<I_{c}$, where $I_{c}$ is the critical current in the
absence of fluctuations \cite{Tinkham,Clarke1,Clarke2}. In this
case, the escape of the Josephson phase can be characterized by
the switching current distribution $P(I_s)$ measured by ramping up
the current at a constant rate. The resonant absorption of photons
in a Josephson junction results in an {\it enhancement} of the
rate of the phase escape at a particular dc bias $I_r$
\cite{Clarke1,Clarke2,Larkin,WalUst}. The value $I_r$ is found
from the resonant condition as the frequency $\omega$ (or $n
\omega$, where $n$ is an integer number, for multiphoton
absorption \cite{WalUst}) matches the energy level separation
\begin{equation}
    \label{ResonantCond1}
    \hbar\omega=E_n(I_r)-E_m(I_r)~~.
\end{equation}
Here, $E_{n,m}(I_r)$ are the quantized energy levels of the
Josephson phase in the presence of an externally applied dc bias
$I_r$. The enhancement of the escape rate manifests itself by
additional microwave-induced peaks in $P(I_s)$. This effect has
been analyzed in detail in Refs.~\cite{Clarke1,Clarke2,Larkin}.

The crucial condition to observe such an enhancement of the escape
rate is the small-signal limit, in which the microwave radiation
only changes the population of individual levels in the well. 
To analyze this condition quantitatively we note that the fluctuation 
induced escape of the Josephson phase from excited levels has to be large. In the
quantum regime the escape probability is governed by the exponent
of the quantum tunneling rate which is proportional to the ratio
of the barrier height to the attempt frequency
$$\frac{E_J (1-\gamma)^{3/2}}{\hbar \omega_p (1-\gamma)^{1/4}}~~,$$
where $E_J$ and $\omega_p$ are the Josephson energy  and the
plasma frequency of the junction, and $\gamma~=~I_s/I_c$. To
observe fluctuation-induced escape of the phase this exponent has
to be of the order of unity, i.e.~$1-\gamma \leq (\hbar
\omega_p/{E_J})^{4/5}$. At the same time to observe the resonance
the drive frequency $\omega$ has to fulfill the resonance
condition $\omega ~=~ \omega_p \left[2(1-\gamma)\right]^{1/4}$.
Thus, an enhancement of the escape that can be directly
interpreted as a microwave-induced increase of the excited level
population, occurs approximately as
\begin{equation} \label{Cond}
\left(\frac{\omega}{\omega_p}\right)^5 \leq \frac{\hbar
\omega_p}{E_J}~~.
\end{equation}

As we turn to a larger microwave frequency or to smaller value of
the critical current density, the condition (\ref{Cond}) is not
valid any more, and a simple increase of the population of excited
levels can {\it not} explain the resonantly induced escape of the
Josephson phase. In this paper we show that in the limit
$({\omega}/{\omega_p})^5~\gg~({\hbar \omega_p})/{E_J}$, the
presence of microwaves leads to an {\it effective suppression} of
the potential barrier for the escape of the phase. This
suppression is particularly strong when the resonance condition is
satisfied. We show that this resonant effect does not only lead to
a double peak structure of the $P(I_s)$ distribution at the
resonance but is also accompanied by a sharp drop in the
dependence of the average switching current $\langle I_s(W)
\rangle$ on the power $W$ of the applied microwave radiation. Note
here, that the analysis of this effect based on the classical
nonlinear dynamics for the Josephson phase has been presented in
\cite{FistUst}.

To analyze the problem quantitatively we write down the
time-dependent potential energy of the Josephson phase $\phi(t)$
in a small Josephson junction in the presence of both dc bias
current $I$ and the ac bias with amplitude $\eta$ induced by
microwaves:
$$
U(\phi) = U_0(\phi) - E_J \frac{\eta}{I_c} \sin(\omega t) \phi~~,
$$
\begin{equation}
\label{UPot} U_0(\phi)~=~-E_J\left(\cos
\phi+\frac{I}{I_{c}}\phi\right)~~~.
\end{equation}
Here,
$U_0(\phi)$ is the potential energy of the Josephson junction in
the absence of ac bias.

In analogy to the classical nonlinear problem \cite{FistUst}, we
write the coordinate $\phi(t)$ as a sum of two terms: a quickly
oscillating resonant term $\xi (t)$ and a term $\phi_0(t)$ slowly
varying in time. Next, the quantum-mechanical average of $\langle
\xi (t) \rangle$ is calculated using perturbation theory with
respect to the small amplitude $\eta$. Moreover, we truncate the
dynamics of the Josephson phase in the potential $U_0(\phi)$ to
the two energy levels $E_{n,m}$ which resonantly interact with the
ac bias. We take into account the damping parameter $\alpha$ that
determines the strength of the resonant interaction. This damping
can be governed by the quasi-particle resistance of the junction
itself and/or dominated by the impedance of the bias leads at the
frequency $\omega$ \cite{Clarke2,Martinis,Makhlin,Fistul}. In linear
approximation we obtain:
\begin{equation} \label{Ksiterm}
\langle \xi(t) \rangle~=~E_J\frac{\eta}{I_c} \, {\rm Im}\left(
\sum_{n,m}
 \frac{f_{nm}^2 e^{i\omega t}}{\hbar^{-1}E_{nm}(I)-\omega +i\alpha}\right),
\end{equation}
where the energy level differences $E_{nm}(I)~=E_n (I)-E_m (I)$
and the matrix elements $f_{nm}~=\langle n|\hat{\phi}|m\rangle$
depend on the dc bias current. The values of $E_n(I)$ and the
corresponding eigenfunctions $| n \rangle $ are found as solutions
of the Schr\"odinger equation
\begin{equation} \label{Schr}
-\frac{\omega_p^2}{2E_J}\Psi_{n}(\phi)^{''}+U_0(\phi)
\Psi_n(\phi) ~=~E\Psi_n (\phi)~~.
\end{equation}

Substituting the $\langle\xi(t)\rangle$ term in the expression for
the potential energy $U(\phi)$ and taking the average over time we
obtain the effective potential energy $U_{\rm{eff}}(\phi_0)$ in
the form
\begin{multline} \label{Ueff}
U_{\rm{eff}}(\phi_0)=-E_J \left[\frac{I}{I_{c}}\phi_0 + \cos
\phi_0 \left(1- \right. \right. \\
\left. \left. \frac{\eta^2}{2}
\sum_{nm}\frac{f_{nm}^4}{(\hbar^{-1}E_{nm}(I)-\omega)^2+\alpha^2
}\right)\right]~~.
\end{multline}

The  dependence of the switching current $I_s(W)$ on the microwave
power $W~=k\eta^2/2$, where $k$ is the microwave coupling
coefficient is obtained by making use of the condition that
$U_{\rm{eff}}(\phi_0)$ has no extreme points. Thus, the shift in
switching current $\delta I_s (W)~= ({I_{c}-I_s(W)})/{I_{c}}$ is
determined by a solution of the transcendental equation
\begin{equation} \label{Treq}
\delta I_s(W)~=~ k^{-1}W\sum_{nm}
\frac{f_{nm}^4}{(\hbar^{-1}E_{nm}(I_s)-\omega)^2+\alpha^2 }~~.
\end{equation}
Note here, that as we derive Eq. (\ref{Treq}) all fluctuation
effects are neglected.

\begin{figure}
\includegraphics[width=0.8\columnwidth]{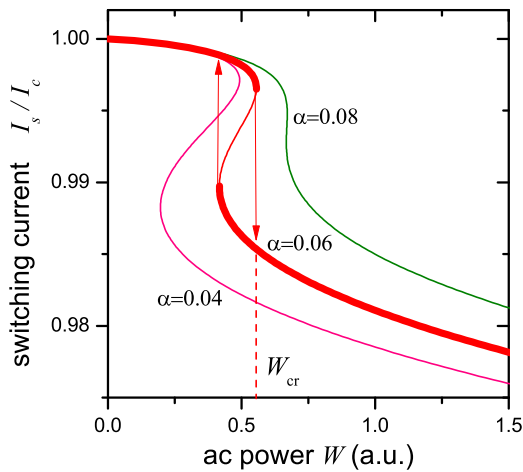}
\caption{ Microwave power dependence of the switching current
$I_s(W)$ of a small Josephson junction. Dashed lines for
$\alpha=0.06$ bound the region in which switching current
distributions with two peaks should be observed. The frequency
$\omega$ of the ac drive is $\frac{\omega}{\omega_p}=0.4$ and the
damping parameter is varied from $\alpha=0.04\,\omega_p$ to
$\alpha=0.08 \,\omega_p$. These correspond to the junction's
effective quality factor $Q=\alpha^{-2}\omega^2$ ranging from 100
to 25.} \label{fig1}
\end{figure}

A good approximation of the dependence $\delta I_s(W)$ can be
obtained in the harmonic oscillator model. In this case, the term
with ($n=0, m=1$) in right-hand part of the Eq.~(\ref{Treq}) is
most important, and the energy level difference is $E_{01}=\hbar
\omega_p (2\delta I_s)^{1/4}$. The typical calculated dependence
of the switching current on the microwave power is presented in
Fig.~\ref{fig1}. The most peculiar feature of the dependence
$I_s(W)$ is a sharp drop as the microwave power is close to the
critical value $W_{cr}$. The appearance of the critical value of
microwave power has a simple physical meaning. As the microwave
power $W$ is less than $W_{cr}$ the resonant interaction is a
weak, and the ac induced suppression of the barrier is not enough
to allow for the Josephson phase to escape. The $I_s(W)$ curves
also display a weak dependence on $W$ in the limit
$W~\leq~W_{cr}$, and this dependence becomes stronger in the limit
of relatively high microwave power $W~\geq~W_{cr}$. Moreover, due
to the multi-valued character of $I_s(W)$ there is a particular
region of microwave power where a switching current distribution
$P(I_s)$ with two peaks should be observed. Notice that, while the
width of the peak corresponding to the larger value of the
switching current $I_s$ is due to the presence of fluctuations
(thermal or quantum), the width of the lower peak is determined by
the damping parameter $\alpha$. The magnitude of the critical
current drop becomes smaller as the damping parameter $\alpha$
increases, see Fig.~\ref{fig1}.

\begin{figure}
\includegraphics[width=0.8\columnwidth]{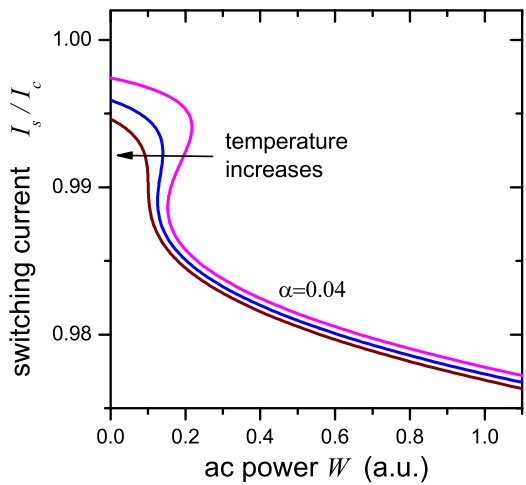}
\caption{Temperature dependence of the switching current of a
small Josephson junction in the presence of ac bias at the
frequency $\omega = 0.4 \, \omega_p$. The reduction of the
critical current due to thermal fluctuations is taken into
account. The damping parameter is fixed $\alpha=0.04\,\omega_p$.
Temperatures shown are 0.1~K, 0.2~K and 0.3~K.} \label{fig2}
\end{figure}

The above analysis can be extended to include fluctuations. The
presence of thermal fluctuations leads to a premature switching of
the junction from the superconducting state with respect to the
value $I_s(W)$ obtained from Eq.~(\ref{Treq}). Taking into account
thermal fluctuations we derive a transcendent equation that is
similar to (\ref{Treq})
$$ \langle\delta I_s(W)\rangle=\langle\delta I_s(0)\rangle+
$$
\begin{equation} \label{Treqfl}
k^{-1}W\sum_{nm} \frac{f_{nm}^4}{(\hbar^{-1}E_{nm}(\langle\delta
I_s(W)\rangle)-\omega)^2+\alpha^2 }~~,
\end{equation}
where $\langle\delta I_s(0)\rangle$ is the fluctuation-induced
shift of the switching current in the absence of ac bias. For
temperatures above the crossover to the quantum regime
$\langle\delta I_s(0)\rangle~=~[\frac{k_B
T}{2E_J}ln(\frac{\omega_p I_c}{2\pi \dot I})]^{2/3}$, where $\dot I
$ is the bias current ramp rate \cite{Tinkham}. By solving the
Eq.~(\ref{Treqfl}) for $\alpha={\rm const}$ we find that thermal
fluctuations lead to smearing of the drop in the mean switching
current and also shift the critical microwave power $W_{cr}$ to
smaller values, as illustrated in Fig.~\ref{fig2}.

Equations (\ref{Treq}) and (\ref{Treqfl}) can also be used in the
more complex case of a strongly non harmonic potential well
$U_0(\phi)$. In such a regime, other terms in Eq.~(\ref{Treq})
become important. For example, by taking into account $E_{02}$
transitions we should expect, according to Eq.~(\ref{Treq}), an
additional jump (of much smaller value) in the dependence $I_s(W)$
\cite{comment}. The microwave power $W$ required to observe this
jump is less than $W_{cr}$, the critical microwave power causing
the main drop in the $I_s(W)$ curve. The multi-photon absorption
of the Josephson phase observed in Ref.~\onlinecite{WalUst}, can
be also be treated in the framework of this analysis. In this case
Eqs.~(\ref{Cond}), (\ref{Treq}) and (\ref{Treqfl}) can still be
used with the substitution of $\omega$ by $n\omega$. Instead of
the linear dependence of $\langle\xi(t)\rangle$ on the amplitude
$\eta$, one should expect the dependence $\langle\xi(t)\rangle\sim
J_n(\frac{\eta}{ 2e \omega})$, where $J_n$ is the Bessel function
of order $n$ \cite{TG}.

The coupling coefficient $k$ can be found by comparison of the
observed $\langle I_s(W)\rangle$ dependence with the one
calculated in Eq.~(\ref{Treqfl}). It allows to calculate the
crucial parameter of coherent quantum-mechanical behavior,
i.e.~the microwave power dependent Rabi frequency of coherent
oscillations as
\begin{equation} \label{RF}
\omega_R=\omega_pf_{nm}\sqrt{\frac{2W}{k}}~~.
\end{equation}

It is worth to note that our analysis is also applicable to the
problem of the microwave-induced escape of a pinned magnetic
fluxon in an annular long Josephson junction, the dynamics of
which can be mapped to the dynamic of a small Josephson junction
\cite{FistUst,WallUstFist2}.

The analysis presented here was motivated by measurements of the
switching current distribution $P(I_s)$, which we performed on a
small ($5 \times 5 \, \rm{\mu m}^2$) Nb/AlO$_x$/Nb Josephson
tunnel junction. This junction had a fluctuation-free critical
current of $I_c = 278.45 \, \rm{\mu A}$, an effective capacitance
of $C = 1.61 \, \rm{pF}$ and a quality factor of $Q \approx 45 \pm
5$ \cite{WalUst}. $P(I_s)$ distributions have been measured at a
bias current ramp rate of $0.245 \, \rm{A/s}$ at the temperature
$T = 100 \, \rm{mK}$, which is well below the temperature
$T^{\star} \approx 280 \, \rm{mK}$, corresponding to the crossover
between thermal escape and quantum tunneling for this sample.
Microwaves in the frequency range between $10 \,\rm{GHz}$ and
$38\,\rm{GHz}$ have been applied to the sample and the microwave
power was varied over a wide range. In Ref.~\onlinecite{WalUst} we
have focused on the regime in which for small microwave powers the
population of an excited state is enhanced and subsequent
tunneling from the excited state is observed. In those
experiments the drive frequency $\omega$ has been chosen to
satisfy the condition (\ref{Cond}). Here, $n$ is the number of
photons involved in the transition. In this case the $P(I_s)$
distributions were only weakly perturbed by the microwaves.

\begin{figure}[tb]
\includegraphics[width=1.0\columnwidth]{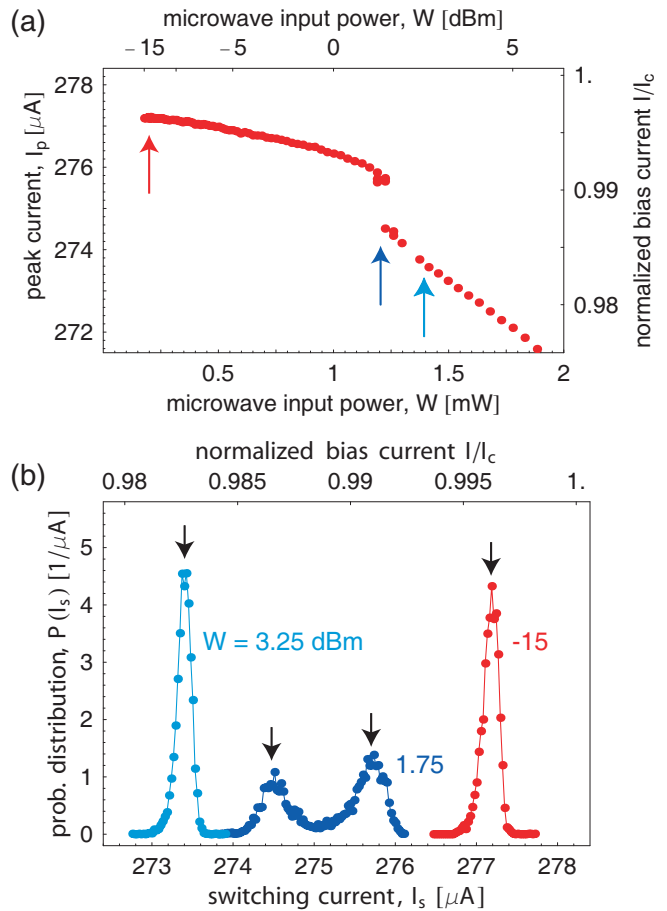}
\caption{(a) Peak switching current $I_p$ versus microwave power
$W$ for a small Josephson junction. The applied microwave
frequency was $\nu=\omega/2\pi = 24\,$GHz ($n \omega / \omega_p =
0.41$). (b) Switching current distributions $P(I_s)$ for the
microwave powers indicated in (a). The peak switching currents
$I_p$ are indicated by arrows.) } \label{fig3}
\end{figure}

Here we present experiments for relatively large microwave
frequencies $n \omega/\omega_p$. To observe a microwave-induced
resonance under this condition, a larger microwave power has to be
applied. In this case, the $P(I_s)$ distributions are strongly
perturbed by the microwaves. In Fig.~\ref{fig3}a the dependence of
the most probable switching current $I_p$ on the microwave power
$W$ is plotted for a microwave frequency of $\nu=\omega/(2\pi)=24
\, \rm{GHz}$. In Fig.~\ref{fig3}b, the $P(I_s)$ distributions are
plotted for the three selected microwave powers indicated in
Fig.~\ref{fig3}a. It is observed that at small microwave powers
the switching current distribution initially shifts to lower bias
currents without developing a resonant double peak structure. At
larger microwave powers -- when the resonance condition is
fulfilled -- a double-peak distribution occurs. It is accompanied
by a pronounced drop in critical current as the microwave power is
further increased. The resonance presented here corresponds to
$n=2$ photons with $2 \omega/\omega_p = 0.41$ \cite{comment2}.

The observed features in the $P(I_s)$ distribution can not be
explained by a simple enhancement of the Josephson phase escape
due to a microwave induced increase of the population of an
excited state \cite{Clarke1,Larkin,WalUst}. We find a particular
microwave power range in which two states (resonant and non
resonant ones) coexist, indicating a microwave-induced bistability
of the junction predicted by our theoretical analysis presented
above. The characteristic behavior in the region of $W~\simeq~W_{cr}$ is
also in a good agreement with analysis above, compare
Fig.~\ref{fig2} and Fig.~\ref{fig3}a \cite{comment3}.

In conclusion we have carried out a quantum-mechanical analysis of
the Josephson phase escape in a small Josephson junction in the
presence of both dc bias and microwave currents. We find that in
the absence of fluctuations and for large microwave frequency the
dependence $I_s(W)$ displays a pronounced drop at resonance and
thereafter decreases smoothly. This behavior is explained by the
effective suppression of the potential barrier by microwaves and
is in a good accord with experiments carried out with small
Josephson junctions. Similar behavior has been observed for a
pinned Josephson vortex in a long annular Josephson junction
\cite{WallUstFist2}.

We acknowledge the partial financial support of this project by
the Deutsche Forschungsgemeinschaft (DFG) and by D-Wave Systems
Inc.


\begin{references}
\bibitem{comment} The crucial condition to observe these additional jumps is
the large values of matrix elements for ac bias induced transitions.
\bibitem{comment2} In order to observe a similar behavior, i.e.~a large
drop in the dependence $\langle I_s(W) \rangle$, for a
single-photon absorption process one would have to apply a larger
microwave frequency $\nu \simeq 50 \, \rm{GHz}$, which is above
the range of our microwave source.
\bibitem{comment3} Notice here that for a two-photon absorption process we have to use 
$\omega~=2(2\pi \nu)$ and replace $W$ by $W^2$ in the Eq. (\ref{Treqfl}).
\end{references}
\end{document}